\documentclass[aps,prl,twocolumn,letterpaper,superscriptaddress,showpacs]{revtex4}
\usepackage{amsmath, amsthm}
\usepackage{keyval}%
\usepackage{graphicx,latexsym}
\usepackage{dcolumn}
\usepackage{bm}
\usepackage{color}
\usepackage{CJK}

\begin{document}
\title{Non-necessity of band inversion process in 2D topological insulators for bulk gapless states and topological phase transitions}

\begin{CJK*}{UTF8}{}

\author{Wenjie Xi (\CJKfamily{gbsn}奚文杰)}
\affiliation{School of Physics and Astronomy, Shanghai Jiao Tong University, Shanghai 200240, China}

\author{Wei Ku (\CJKfamily{bsmi}顧威)}
\altaffiliation{corresponding email: weiku@mailaps.org}
\affiliation{Tsung-Dao Lee Institute, Shanghai 200240, China}
\affiliation{School of Physics and Astronomy, Shanghai Jiao Tong University, Shanghai 200240, China}
\affiliation{Key Laboratory of Artificial Structures and Quantum Control (Ministry of Education), Shanghai 200240, China}

\date{\today}

\begin{abstract}
In commonly employed models for 2D topological insulators, bulk gapless states are well known to form at the band inversion points where the degeneracy of the states is protected by symmetries.
It is thus sometimes quite tempting to consider this feature, the occurrence of gapless states, a result of the band inversion process under protection of the symmetries.
Similarly, the band inversion process might even be perceived as necessary to induce 2D topological phase transitions.
To clarify these misleading perspectives, we propose a simple model with a flexible Chern number to demonstrate that the bulk gapless states emerge at the phase boundary of topological phase transitions, despite the absence of band inversion process.
Furthermore, the bulk gapless states do not need to occur at the special $k$-points protected by symmetries.
Given the significance of these fundamental \textit{conceptual} issues and their wide-spread influence, our clarification should generate strong general interests and significant impacts.
Furthermore, the simplicity and flexibility of our general model with an arbitrary Chern number should prove useful in a wide range of future studies of topological states of matter.

\end{abstract}

\maketitle
\end{CJK*}

The topological materials have raised great attention in the past decade because of their intrinsic physical importance and potential application in spintronics~\cite{Paper1,Paper2} and quantum computation~\cite{Paper3,Paper4}.
As the first discovered topological insulators~\cite{Paper14,Berry,Paper5,Theory,Paper11,Paper12,Paper6,Paper7,Paper8,Paper9,Yu61,Paper10,Paper13,Von,Hsieh2008,Xia2009,Ivan2011,Jia2012,Du2015,Mb2,Mb3,Exp,Inv,Niu2016}, the two-dimensional topological insulators have been intensively studied both theoretically~\cite{Paper14,Berry,Theory,Paper5,Paper11,Paper12} and experimentally~\cite{Paper10,Von,Hsieh2008,Xia2009,Ivan2011,Jia2012,Du2015,Paper13}.
They can be characterized by the topological invariants, like the Chern number~\cite{Paper14,Paper15} defined initially only for two-dimensional insulators.
Recently, with further discovery of the three-dimensional analogues of topological insulators, the abundant nontrivial phases of 3D topological materials have continued to draw much attention of the community.
Some of these striking topological materials are, for example, Weyl semimetals~\cite{Paper21,Paper16,Paper17,Paper18,Paper19,Paper20}, topological nodal line semimetals~\cite{Paper23,Paper22} and Dirac semimetals~\cite{Paper25,Paper24}.
Currently, researches are also focusing on topological superconductors~\cite{Paper27,Paper26} whose gapless states at the edge give the long sought Majorana bound states~\cite{Paper28}.

\begin{figure*}[th]
\begin{center}
\resizebox*{2.0\columnwidth}{!}{\includegraphics{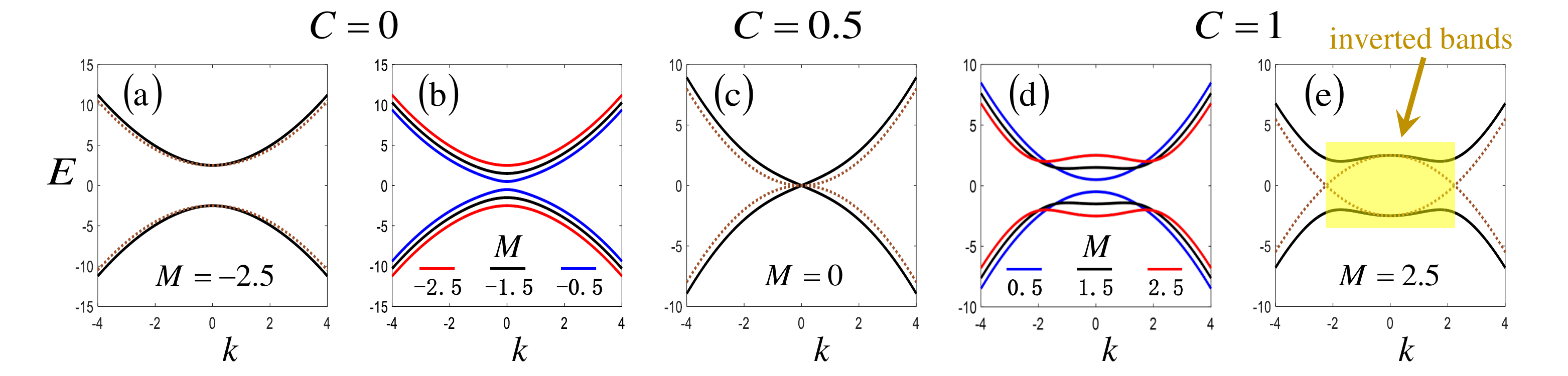}}
\end{center}
\vspace*{-0.8cm}
\caption{
\label{fig:fig1}
Illustration of occurrence of the bulk gapless states during the band inversion process in typical models.
The dispersion of $d_3(k)$ (dashed lines) makes clear that as $M$ moves from negative (a) through zero (c) to positive (e), the bands become inverted in the yellow region near $k=0$.
While the gap opens up upon introduction of the off-diagonal coupling $d_1$ and $d_2$ in (b) and (d), bulk gapless states are preserved at $M=0$ (c) due to \textit{a priori} absence of coupling at $k=0$ in the models (presumably protected by some unspecified symmetry).
}
\vspace*{-0.5cm}
\end{figure*}

Specifically for 2D topological insulators, bulk gapless states also appear in the typical models~\cite{Paper5,Paper29} at special $k$-points, where couplings between bands are forbidden by symmetries.
In the absence of such couplings, the band will thus inevitably close the gap and produce bulk gapless states at these $k$-points during the band inversion process, as demonstrated in Fig.~\ref{fig:fig1}.
(Here the band inversion process corresponds to change of the sequence of the eigen-vectors in some region of the crystal momentum $k$.)
It is thus quite tempting to associate the occurrence of the bulk gapless states (or perhaps even the topological phase transitions) with the band inversion process.
In fact, many articles in the current literature could be perceived as rather misleading on these important conceptual issues~\cite{Paper32,Paper33,Paper36,Paper37,Mb5,Paper40,Paper41,Paper42,Paper43,Paper44,Paper45} .

In this letter, we demonstrate that the essential aspect of occurrence of the bulk gapless states is only the topological phase transition, and that the band inversion process is conceptually not necessary, even though it might be common in real materials.
We first examine the usual considerations using a widely applied model for 2D topological insulators.
We then propose a simple model that allows an arbitrary Chern number and use it to demonstrate that topological phase transitions can take place without ever involving band inversion process.
Finally, we illustrate that bulk gapless states always appear in the phase boundary even without band inversion process, and their location do not need to be protected by any symmetry \emph{a priori}.
This study not only clarifies common conceptual issues of great significance, but also proposes a simple, flexible, and general model for future investigations of topological phase transitions in topological systems.

As a representative example, let's first review the BHZ model~\cite{Paper29} for 2D topological insulators, expressed by a two-band Hamiltonian in the 2D crystal momentum $k$- and orbital $m$-space:
\begin{eqnarray}
\label{eq:eqn1}
&&\langle\vec{k},m|\hat{H}|\vec{k},m^\prime\rangle\rightarrow\nonumber\\
&&H(\vec{k})=\vec{d}(\vec{k})\cdot \vec{\sigma}=
\left(
\begin{array}{cc}
d_3 & d_1-id_2\\
d_1+id_2 & -d_3
\end{array}
\right),
\end{eqnarray}
where $d_1=k_x$, $d_2=k_y$, $d_3=M-|\vec{k}|^2/2$ with a tunable mass parameter $M$.
Expressed in $\vec{b}=\vec{d}/d$,$d=|\vec{d}|$, and introduce a third axis $z$ for simplier 3D notation $\nabla=(\partial_{k_x}, \partial_{k_y}, \partial_z)$, we have the Berry connection $\vec{A}(\vec{k})=i\langle\vec{k},j_0|\nabla|\vec{k},j_0\rangle$ and the Berry curvature $\vec{B}(\vec{k})=\nabla\times\vec{A}=\frac 12 (b_1\nabla b_2\times \nabla b_3+b_2\nabla b_3\times \nabla b_1+b_3\nabla b_1\times \nabla b_2)$, where $|\vec{k},j_0\rangle$ is the eigenvector of the lower band of $H$.
This model goes through a topological phase transition via a band inversion process at $M = 0$.
As shown in Fig.~\ref{fig:fig1}, before the band inversion, $M < 0$, the system is topologically trivial with a zero Chern number $C=\frac{1}{2\pi}\int_{k_x k_y} \vec{B}\cdot \mathrm{d} \vec{S}=0$.
After the band inversion, $M > 0$, the sequence of the eigenvectors is inverted near $k = 0$ [within the yellow region of Fig.~\ref{fig:fig1}(e)], as made clear from the band crossing in the dispersion of $d_3$ (dashed lines), and the system becomes topologically nontrivial with $C = 1$.
Notice that the band gap systematically reduces, closes, and reopens as the band inversion takes place, and corresponding bulk gapless states occur right at the band inversion point, $M = 0$.

\begin{figure*}[th]
\begin{center}
\resizebox*{2.0\columnwidth}{!}{\includegraphics{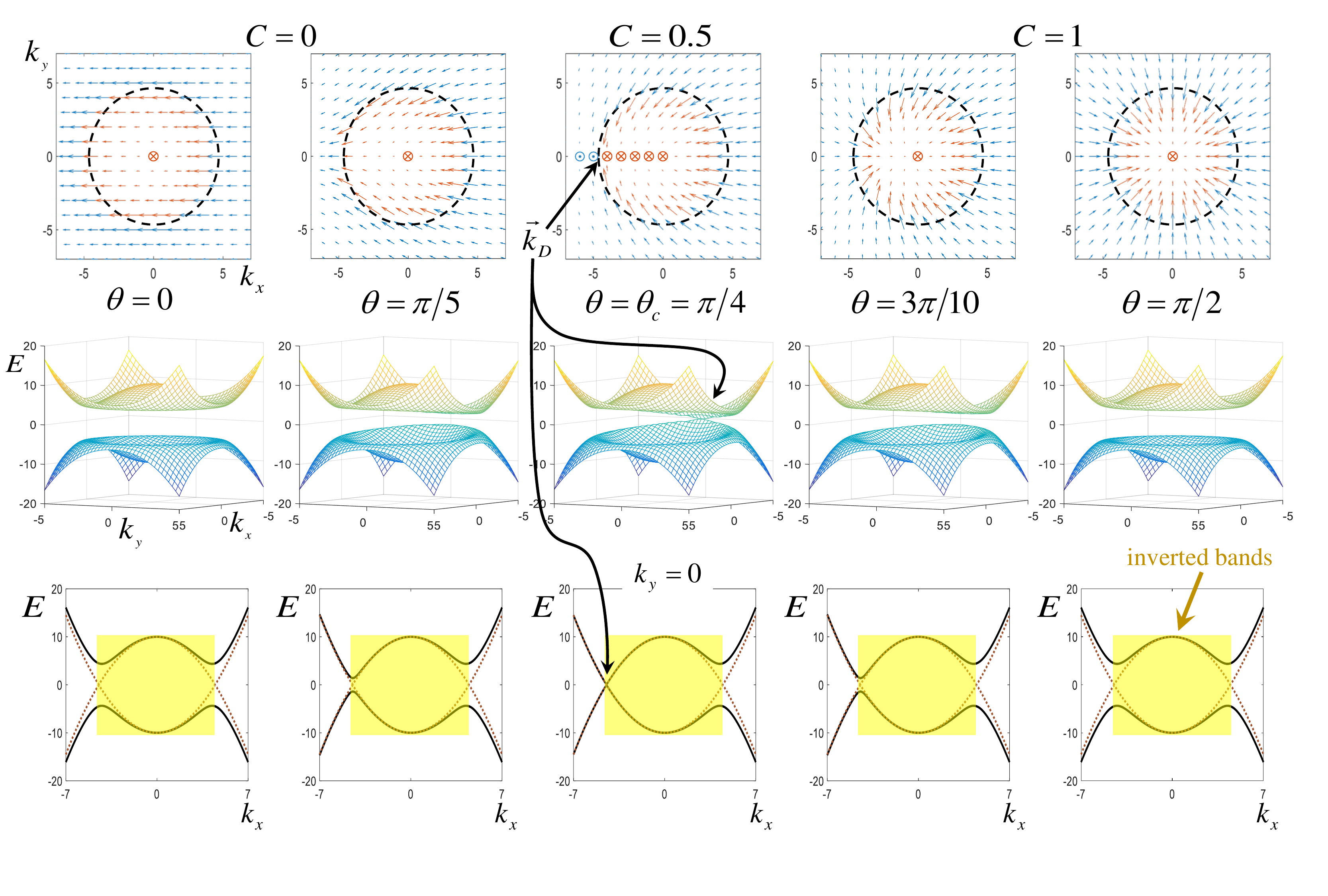}}
\end{center}
\vspace*{-1.2cm}
\caption{
\label{fig:fig2}
Illustration of pseudo-spins (upper panels) and band dispersion in the 2D $k$-space (middle panels) and along the ($k_x$,0) path (lower panels), across the $(n,n^\prime)=(0,1)$ topological phase transition from $C=0$ (two left panels) to $C=1$ (two right panels) across the phase boundary at $\theta_c$ (center panels).
No band inversion process takes place, since with a fixed $M=10$ the bands remain inverted inside the $d_3=0$ contour (dashed lines in the upper panels and the yellow region in the lower panels.)
Blue (red) arrows represent the $xy$-components of the pseudo-spins having positive (negative) $z$-components.
At the phase boundary $\theta=\theta_c$, $\vec{k}_D$ denotes the location of the non-analytic $\pi$-jumps of pseudo-spins, where the bulk gapless states appear.
}
\vspace*{-0.5cm}
\end{figure*}

In this model, the occurrence of bulk gapless states is easily observed from the structure of the matrix near $k = 0$ at $M = 0$, which contains degenerate diagonal elements $\pm d_3$ and zero off-diagonal elements $d_1\pm id_2$ that supposedly results from underlying symmetries of the system.
This common perspective, however, tends to miss the profound principle that gapless states are inevitable in the boundary of topological phase transitions, regardless of the band inversion process or any symmetry protection of the gapless point.

To illustrate conceptually the most fundamental nature of the bulk gapless states, we demonstrate below their occurrence across the topological phase transition without a band inversion process.
Furthermore, to omit unnecessary microscopic details of real materials and to prevent unintentional introduction of symmetry-related (but topologically irrelevant) features, we focus on the low-energy effective Hamiltonian near the divergent point of the Berry curvature.
The common considerations of symmetries thus only serve to reduce various microscopic systems to the general effective picture, and will not play any relevant role in the following analysis.
To this end, we extend the above BHZ model to allow arbitrary Chern number $C = n$:
\begin{eqnarray}
\label{eq:eqn2}
&&\langle\vec{k},m|\hat{H}|\vec{k},m^\prime\rangle\rightarrow\nonumber\\
&&H^{(n)}(\vec{k}) \equiv \vec{d}^{(n)}(\vec{k})\cdot \vec{\sigma}=
\left(
\begin{array}{cc}
d_3 & d_0 e^{-in\phi}\\
d_0 e^{in\phi} & -d_3
\end{array}
\right),
\end{eqnarray}
where $d_0=k,d_3=M-\frac 12 k^2$, and $\vec{k}=(k,\phi,z)$ in the cylindrical coordinate system.
It is straightforward to show that, 
\begin{eqnarray}
\label{eq:eqn3}
\vec{A}&=&\hat{\phi} \frac{n}{2k}(1-b_3) + \hat{k} f(k),\\
\vec{B}&=&-\hat{z}\frac{n}{2k}\partial_k b_3,\\
C &=& -\left. \frac n2b_3 \right|_0^\infty=
\left\{
\begin{array}{c}
	n, M>0\\
	\frac n2, M=0\\
	0, M<0
\end{array},
\right.
\end{eqnarray}
where $f(k)$ is independent of $\phi$ and thus does not contribute to $\vec{B}$.
This model reproduces to the above BHZ model at $n = 1$, and similar to the BHZ model, it allows topological phase transitions from $C = 0$ to $C=n$ by varying $M$ across zero with a fixed $n$.
Alternatively, it also allows topological phase transitions from $C = n$ to $C=n'$ for a fixed $M > 0$,
\begin{eqnarray}
\label{eq:eqn4}
H^{(n,n^\prime)}(\vec{k})&\equiv&\vec{d}^{(n,n^\prime)}(\vec{k})\cdot \vec{\sigma}\\
&=&[\cos^2(\theta)\vec{d}^{(n)}(\vec{k})+\sin^2(\theta)\vec{d}^{(n^\prime)}(\vec{k})]\cdot \vec{\sigma}\nonumber
\end{eqnarray}
by tuning parameter $\theta$ from $0$ to $\pi/2$.
The transition now takes place at $\theta_c=\pi/4$, and does not involve a band inversion process, since the bands are kept partially inverted at $M > 0$.

We first examine the case of $(n, n') = (0, 1)$ with $M=10$ shown in Fig.~\ref{fig:fig2}.
At $\theta<\theta_c$, the system is in a $C=0$ topologically trivial phase, even though the bands are inverted near $k=0$, evident from the yellow region in the lower panels, in which the dashed bands of $d_3$ cross each other.
The inverted bands can also be observed from the change of $z$-component of the corresponding pseudo-spins $-\vec{d}^{(n,n^\prime)}(\vec{k})/|\vec{d}^{(n,n^\prime)}(\vec{k})|$ from blue (positive) to red (negative) inside the dashed lines in the upper panels as well.
(A general symmetry-independent, coordinate-invariant definition of inverted bands can be made by identifying a contour in the 2D $k$-space, dashed lines in the upper panels of Fig.~\ref{fig:fig2}, on which the pseudo-spin resides only in a plane, and across the contour the perpendicular component changes sign.)
Even with inverted bands, the phase is nonetheless topologically trivial corresponding to a zero net rotation of the pseudo-spins along the $d_3=0$ contours (dashed lines) in the first two panels.
On the other hand, at $\theta>\theta_c$, one observes a complete counter-clockwise $2\pi$ rotation of the pseudo-spins along the $d_3=0$ contours in the last two panels, reflecting the topological invariant $C=1$.
Most interestingly, at the phase boundary $\theta=\theta_c$ (the center panels), the pseudo-spins develop a net counter-clockwise rotation of $\pi$ along the contour and a non-analytic $\pi$-jump at $\vec{k}_D=(k,\phi,z)=(\sqrt{2M}, \pi, 0)$, across which the pseudo-spins are opposite in directions.

\begin{figure*}[th]
\begin{center}
\resizebox*{2.0\columnwidth}{!}{\includegraphics{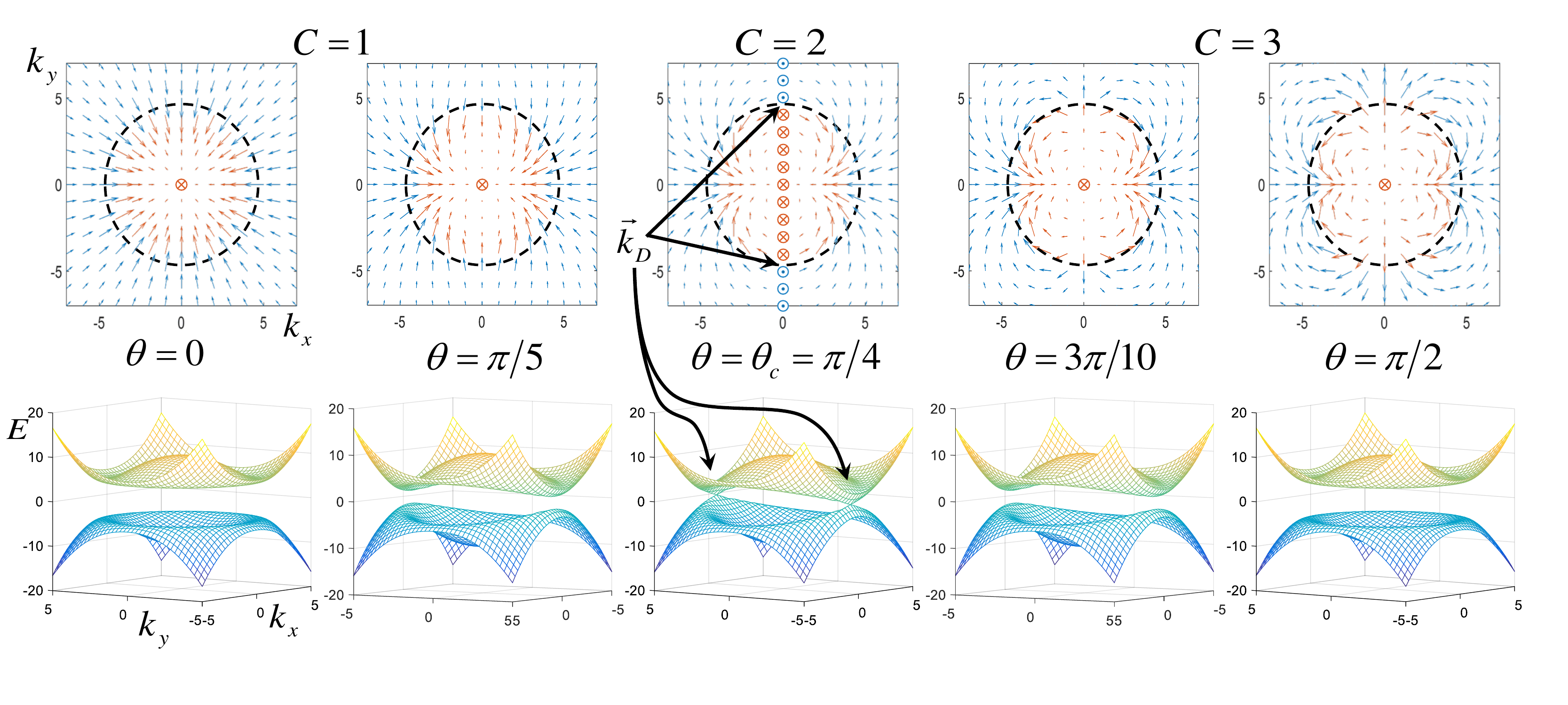}}
\end{center}
\vspace*{-1.2cm}
\caption{
\label{fig:fig3}
Same as Fig.~\ref{fig:fig2} but with $(n,n^\prime)=(1,3)$
}
\vspace*{-0.5cm}
\end{figure*}

Now, even without the band inversion process, the middle and lower panels of Fig.~\ref{fig:fig2} show clearly that across the topological phase transition, the gap still closes and reopens, with bulk gapless states emerging at $\vec{k}_D$ right at the phase boundary.
Furthermore, the location of the bulk gapless states $\vec{k}_D$ is not \emph{a priori} protected by any symmetry, since the off-diagonal elements are not zero when $\theta\ne\theta_c$.
This example illustrates clearly that the occurrence of the gapless states is a necessary outcome of the topological phase transitions, but not necessarily related to the band inversion process or any symmetry protection of the corresponding $k$-points.
It should also be obvious that topological phase transitions do not necessarily require the band inversion process either.

These conclusions can be verified again by examining the case of $(n, n') = (1, 3)$ with $M=10$.
The last two panels of Fig.~\ref{fig:fig3} show the similar behavior of $3\times 2\pi$ counter-clockwise rotation of the pseudo-spins along the $d_z=0$ contours at $\theta>\theta_c$, corresponding to $C=3$.
Similar to the previous case, at the phase boundary $\theta=\theta_c$, the pseudo-spins host non-analytic $\pi$-jumps, now at $\vec{k}_D=(k,\phi,z)=(\sqrt{2M},\pm\pi/2,0)$ that satisfies in general $\cos[(n-n^\prime)\phi] + 1 = 0$.
Such $\pi$-jumps each introduces a $C=\pm0.5$ change in comparison to the neighboring topological phases, and is thus responsible for changing $C$ by one across the phase boundary.
Correspondingly, the third panel of Fig.~\ref{fig:fig3} shows bulk gapless states appearing at these two $\vec{k}_D$ points.
This case again emphasizes the non-necessity of the band inversion process for the occurrence of the bulk gapless states (or of topological phase transitions.)

We note that our main emphasis is to clarify the conceptually fundamental issues, even though our model might not seem to have a direct connection to a known 2D topological insulating material at the moment.
Nonetheless, our conclusion of non-necessity of the band inversion process is not merely conceptual, but a realistic consideration consistent with the recent proposal of topological states of non-Dirac electrons on a triangular lattice~\cite{Hu2016}, as in MoS$_2$ monolayers endowed with 3d transition metal adatoms~\cite{Without1}.

In summary, we clarify a common misconception that the occurrence of bulk gapless states (and perhaps even the topological phase transitions) is associated with the band inversion process.
To this end, we propose a model for 2D topological insulators that allows any Chern number.
We then demonstrate that gapless states are bound to take place during topological phase transitions, and neither of them requires the band inversion process.
Furthermore, a symmetry protection against coupling between bands is unnecessary at the $k$-points of the bulk gapless states.
Given the significance of these fundamental \textit{conceptual} issues and their wide-spread influence, our clarification should generate strong general interests and significant impacts.
Furthermore, the simplicity and flexibility of our general model with an arbitrary Chern number should prove useful in a wide range of future studies of topological states of matter.

Work supported by National Natural Science Foundation of China \#11674220 and 11447601, and Ministry of Science and Technology \#2016YFA0300500 and 2016YFA0300501.

\bibliography{Ti}

\end{document}